# RadImageGAN – A Multi-modal Dataset-Scale Generative AI for Medical Imaging

[1]Zelong Liu, [1]Alexander Zhou, [1]Arnold Yang, [1]Alara Yilmaz, [1]Maxwell Yoo, [1]Mikey Sullivan, [1]Catherine Zhang, [1]James Grant, [2]Daiqing Li, [1,5]Zahi A. Fayad, [2,5]Sean Huver, [3,4,5]Timothy Deyer, [1,5]Xueyan Mei

[1]BioMedical Engineering and Imaging Institute, Icahn School of Medicine at Mount Sinai, New York, NY

[2]NVIDIA, Santa Clara, CA

[3]East River Medical Imaging, New York, NY

[4]Department of Radiology, Cornell Medicine, New York, NY

[5]These authors jointly supervised this work.

Corresponding authors: Zahi Fayad, email: zahi.fayad@mssm.edu; Sean Huver, email: shuver@nvidia.com; Timothy Deyer, email: tdeyer@eastriverimaging.com; Xueyan Mei, email: xueyan.mei@icahn.mssm.edu.

## ABSTRACT

Deep learning in medical imaging often requires large-scale, high-quality data or initiation with suitably pre-trained weights. However, medical datasets are limited by data availability, domain-specific knowledge, and privacy concerns, and the creation of large and diverse radiologic databases like RadImageNet is highly resource-intensive. To address these limitations, we introduce RadImageGAN, the first multi-modal radiologic data generator, which was developed by training StyleGAN-XL on the real RadImageNet dataset of 102,774 patients. RadImageGAN can generate high-resolution synthetic medical imaging datasets across 12 anatomical regions and 130 pathological classes in 3 modalities. Furthermore, we demonstrate that RadImageGAN generators can be utilized with BigDatasetGAN to generate multi-class pixel-wise annotated paired synthetic images and masks for diverse downstream segmentation tasks with minimal manual annotation.

We showed that using synthetic auto-labeled data from RadImageGAN can significantly improve performance on four diverse downstream segmentation datasets by augmenting real training data and/or developing pre-trained weights for fine-tuning. This shows that RadImageGAN combined with BigDatasetGAN can improve model performance and address data scarcity while reducing the resources needed for annotations for segmentation tasks.

## KEY POINTS

1. The RadImageGAN generator can produce high-quality multi-class/modality synthetic images across 130 pathologies and 12 anatomies in CT, MRI, and endoscopy imaging modalities
2. BigDatasetGAN can be used with RadImageGAN to generate paired masks for fully labeled multi-class synthetic images with a weakly supervised approach requiring minimal manual annotation
3. Using synthetic labeled imaging data to develop pre-trained weights for transfer learning ("synthetic pretraining") and/or augmenting real medical imaging datasets in training ("synthetic augmentation") can significantly boost performance on downstream segmentation applications, particularly when the downstream datasets are small

# INTRODUCTION

Medical imaging has revolutionized the field of healthcare over the past century by providing accurate and non-invasive diagnostic tools. Within the last ten years, machine learning has undergone a parallel revolution in healthcare, providing diagnostic, predictive, and treatment models for medical imaging leading to increased diagnostic accuracy, improved treatment effectiveness, and cost efficiencies (1,2). To reach these performance standards, models require large representative volumes of high-quality medical imaging data (1,3). However, the data collection and manual annotation of medical images for training these models is a time-consuming and resource-intensive task requiring expert medical domain knowledge, which limits the size and diversity of available datasets for machine learning applications (4). Furthermore, the publication and sharing of medical datasets are limited by significant privacy and regulatory concerns (5–9), resulting in datasets that are limited in size and scope, data siloing within health systems, and unbalanced representation of general populations or other rarer conditions. Addressing these issues will unlock the full potential of machine learning in developing novel medical applications and delivering enhanced healthcare.

Recently, RadImageNet was curated as a large-scale radiological image database that contains 1.35 million images in 3 modalities (CT, MRI, and US) labeled with 165 pathologic classes on 11 anatomic regions, making it a valuable resource for training deep learning models (10). However, the creation of such a database was resource-intensive, requiring significant amounts of time and expertise to collect and annotate images.

Traditional methods to circumvent these data limitations include transfer learning from larger and/or out-of-domain datasets and basic data augmentation on existing image data (e.g., geometric transformations, random erasing, image mixing, kernel filtering). However, these methods do not solve the fundamental issue of data availability. More recently, generative machine learning approaches are being studied to enable the creation of synthetic medical imaging datasets with similar aggregate statistical characteristics to real source datasets for various applications in medical imaging. A promising paradigm that has emerged is the use of generative adversarial networks (GANs) (11) to create synthetic medical images for the training of machine learning models (12–17).

Gao et al. developed SctheX, a framework for developing generalizable AI algorithms from annotated computed tomography (CT), to generate synthetic X-ray images for model training and achieved comparable or better performance than CycleGAN (18) and other models trained directly on real data in three downstream clinical tasks (hip imaging segmentation, surgical robotic tool detection, and COVID-19 lesion segmentation) (19). Guo et al. proposed MedGAN, a GAN architecture using Wasserstein loss as a convergence metric, which was used to generate medical images for few-shot learning models on various disease classification and lesion localization datasets (15). Gheorghiță et al. used a GauGAN model (20) to generate synthetic short-axis cardiac Cine magnetic resonance imaging (MRI) scans for convolutional neural network (CNN) pretraining, which significantly improved cardiac function quantification (21). Other work demonstrated the use of GANs for synthetic data augmentation in liver lesion classification (14) and generation of indistinguishable high-resolution chest radiograph synthesis (22), aggregated GANs for brain tumor image generation (23), and pix2pix-based GAN for lung cancer computed tomography (CT) images (24).

However, these approaches are limited, as these generative architectures are typically restricted to synthesizing imaging data in a specific modality, anatomy, and/or pathology. Furthermore, the generated imaging data may lack semantic segmentation, which could provide further downstream utility.

In this study, we propose training StyleGAN-XL, a new state-of-the-art GAN model that scales StyleGAN for multi-class large unstructured datasets (25), on RadImageNet to generate synthetic radiological images that fall within the diagnostic classes of RadImageNet. With this approach, we aim to develop a multi-class radiologic data generator, which we term "**RadImageGAN**", that can generate synthetic labeled multi-class medical images. In this study, we specifically develop 2 generators: "**RadImageGAN-CT/MR**", which generates CT and MR images with 124 pathological classes created from the real data of 102,774 patients, and "**RadImageGAN-Gastro**", which generates gastrointestinal colonoscopy video images with 6 pathological classes, for a total of 130 classes.

To further enhance the utility of RadImageGAN, we then aim to apply BigDatasetGAN (26) to RadImageGAN to develop a multi-class, pixel-wise label generator of RadImageGAN synthetic images, "**RadImageGAN-Labeled**". BigDatasetGAN is a GAN-based approach that can generate pixel-wise labeled synthetic images with a small set of manually labeled GAN-generated images. Applying BigDatasetGAN to RadImageGAN allows us to generate multi-class synthetic and fully annotated images with minimal manual labeling effort, which can then be used to create large-scale datasets for training segmentation models and augment downstream segmentation datasets.

The approach and utility of this new paradigm is illustrated in **Figure 1**. In this study, we demonstrate the ability of RadImageGAN to generate high-quality multi-class synthetic CT and MR imaging and its ability to boost performance in medical imaging segmentation tasks in diverse data availability conditions (low, moderate, and full real data availability) when combined with BigDatasetGAN. Improved performance comes from using RadImageGAN for: 1) synthetic data augmentation of real datasets for model training ("**synthetic augmentation**") and 2) development of pre-trained weights from synthetic data for downstream transfer learning ("**synthetic pretraining**"). We find that our RadImageGAN-generated synthetic labeled images and pre-trained weights can significantly boost segmentation performance compared to models solely trained on real data under most data availability conditions.

**RESULTS**

*StyleGAN-XL training for RadImageGAN*

StyleGAN-XL models can generate high-resolution synthetic images by learning from large and diverse unstructured datasets. This new state-of-the-art approach enables our vision to generate large numbers of high-resolution synthetic medical images across pathologies, modalities, and anatomies. Using 8 NVIDIA DGX-A100 GPUs with 640GB VRAM, the RadImageGAN-CT/MR model was trained on the 0.88 million CT/MR images and 124 corresponding unique pathological labels from the RadImageNet database. **Table S2** summarizes the training progress of RadImageGAN-CT/MR from low resolution (16x16 px) to high resolution (512x512 px) image output over 4,563 A100 computational hours. The Fréchet inception distance (FID) score

was 6.40, 7.75, 7.60, 8.85, 9.23, and 9.37 at the output resolution stages of 16x16, 32x32, 64x64, 128x128, 256x256, and 512x512 px, respectively.

The RadImageGAN-Gastro model was developed on a subset of the HyperKvasir gastrointestinal dataset consisting of 6 pathological labels and 5,713 images from colonoscopy procedures (27). **Table S3** summarizes the training progress of RadImageGAN-Gastro on 8 NVIDIA DGX-V100 GPUs with 256GB VRAM over 3,088 computational hours. The FID score was 8.08, 9.12, 11.36, 8.74, 5.50, and 5.05 at the output resolution stages of 16x16, 32x32, 64x64, 128x128, 256x256, and 512x512 px, respectively.

**Figures 2** and **3**, and **Table S1** provide illustrative examples and summarize the modalities, views, anatomies, and labels that RadImageGAN is able to generate. Generator training time and performance are summarized in **Tables S2** and **S3**.

*Training and performance of RadImageGAN-Labeled on downstream applications*

To assess the utility of RadImageGAN-CT/MR and RadImageGAN-Gastro in diverse downstream medical applications, the following four public segmentation datasets and tasks were selected for their diversity in anatomy, modality, and pathology: BTCV-Abdomen (abdominal CT) (28), CHAOS-MRI (abdominal MRI) (29), Labeled Lumbar Spine MRI (spinal MRI) (30), and CVC-ClinicDB (gastrointestinal endoscopy images) (31). For BTCV-Abdomen, the normal abdomen CT class from RadImageGAN-CT/MR was used for synthetic generation, and BigDatasetGAN was then used to develop an automatic mask generator for liver and kidney segmentation. For CHAOS-MRI, the normal abdomen MRI class from RadImageGAN-CT/MR was used for synthetic generation, and BigDatasetGAN was then used to develop an automatic mask generator for liver and kidney segmentation. For Labeled Lumbar Spine MRI, the spinal canal stenosis MRI class from RadImageGAN-CT/MR was used for synthetic generation, and BigDatasetGAN was then used to develop an automatic mask generator for intervertebral disc (IVD), posterior element (PE), and thecal sac (TS) segmentation. For CVC-ClinicDB, synthetic images of the polyp colonoscopy imaging class from RadImageGAN-Gastro were used for synthetic generation, and BigDatasetGAN was then used to develop an automatic mask generator for polyp segmentation. Downstream tasks are summarized in **Table S8**.

For each downstream task, the efficacy of the synthetic images for data augmentation in model training ("**synthetic augmentation**") was assessed by training nnU-Net models (32) under 10 data conditions for each of the 4 datasets with varying proportions of real and synthetic imaging data. The data conditions are summarized in **Table 1**. The BTCV-Abdomen and CHAOS-MRI datasets collected data by patient scans, so those minimal and low real data conditions used real imaging data from 1 and 2 full patient scans, respectively. The Labeled Lumbar Spine MRI and CVC-ClinicDB datasets collected data at a slice level, so those minimal and low real data conditions used 1% and 10% of total available real slice images, respectively. Synthetic augmentation was then used to add synthetic images such that the augmented training dataset was approximately the same size as the full real training dataset, which we call "synthetic gap augmentation." For full real data conditions, we used the full real training datasets and augmented them with different amounts of synthetic images (10%, 50%, and 100% of the full real training dataset size).

The efficacy of these synthetic images for transfer learning ("**synthetic pretraining**") was also assessed by pretraining a nnU-Net model for each of the 4 datasets on a corresponding synthetic dataset similarly sized to each of the real training datasets and subsequently fine-tuning each of the 4 pre-trained models under the same 10 data conditions. This assessed the effects of fine-tuning synthetically pretrained models with both purely real and augmented datasets.

Performance was quantified by the average Dice score of resulting segmentation on test datasets in all experiments. A summary of the relative effects of the two synthetic methods in all experiments is shown in **Figure 4**.

*BTCV-Abdomen*

The BTCV-Abdomen dataset consists of 30 abdominal CT patient scans. For each fold of the five-fold cross-validation, the minimal/low/moderate/full real training data condition was defined as 1/2/11(50%)/22(100%) patient abdominal scans; 2 scans were used for validation, and 6 scans were used for testing. Results are summarized in **Table 2a,** and corresponding statistics in **Tables S4a** and **S4b**.

For the BTCV-Abdomen dataset, training nnU-Net models in minimal real data conditions (1 scan) achieved an average Dice score of 0.582/0.482 on liver/kidney segmentation. Synthetic gap augmentation in training boosted performance to 0.678/0.645 on liver/kidney segmentation ($p<0.001/<0.001$). Using a pre-trained model and fine-tuning on minimal real data boosted performance to 0.636/0.590 ($p<0.001/p<0.001$), while using a pre-trained model and fine-tuning on minimal real data with synthetic gap augmentation yielded similar results (0.677/0.654; $p<0.001/p<0.001$) to solely training on minimal real data with synthetic gap augmentation without synthetic pretraining.

Training in low real data conditions (2 scans) achieved an average Dice score of 0.648/0.582 on liver/kidney segmentation. Synthetic gap augmentation in training boosted performance to 0.713/0.675 on liver/kidney segmentation ($p=0.971/p<0.001$). Using a pre-trained model and fine-tuning on low real data boosted performance to 0.698/0.677 ($p<0.001/p<0.001$), while using a pre-trained model and fine-tuning on low real data with synthetic gap augmentation yielded similar results (0.702/0.673; $p=0.468/p<0.001$) to solely training on low real data with synthetic gap augmentation without synthetic pretraining.

Training on moderate real data conditions (11 scans) achieved an average Dice score of 0.823/0.813 on liver/kidney segmentation. Synthetic gap augmentation mildly boosted performance to 0.829/0.825 ($p<0.001/p<0.001$). Using a pre-trained model and fine-tuning on moderate real data boosted performance to 0.846/0.833 ($p<0.001/p<0.001$), while using a pre-trained model and fine-tuning on moderate data with synthetic gap augmentation yielded similar results (0.837/0.823; $p<0.001/p<0.001$) to solely training on moderate data with synthetic gap augmentation without synthetic pretraining.

Training on full real data conditions (all 22 scans) achieved an average Dice score of 0.833/0.832 on liver/kidney segmentation. Augmenting with 10%, 50%, and 100% synthetic data yielded similar performances of 0.834/0.856, 0.845/0.854, and 0.841/0.847, respectively ($p<0.001/p=0.710$, $p<0.001/p=0.061$, $p<0.001/p<0.01$). Using a pre-trained model and fine-tuning on full real data boosted performance to 0.852/0.863 ($p<0.001/p<0.001$), while using a pre-trained model and fine-tuning on full real data with 10%, 50%, and 100% synthetic data augmentation yielded similar results of 0.847/0.865, 0.841/0.849, and 0.847/0.850, respectively (p=0.104/p=0.133, p<0.001/p=0.379, p<0.001/p<0.01).

*CHAOS-MRI*

The CHAOS-MRI dataset consists of 20 T2 abdominal MRI patient scans. For each fold of the five-fold cross-validation, the minimal/low/moderate/full real data condition was defined as 1/2/7(50%)/14(100%) patient abdominal scans; 2 scans were used for validation, and 4 scans were used for testing. Results are summarized in **Table 2b,** and corresponding statistics in **Tables S5a** and **S5b**.

For the CHAOS-MRI dataset, training nnU-Net models in minimal real data conditions (1 scan) achieved an average Dice score of 0.446/0.394 on liver/kidney segmentation. Synthetic gap augmentation in training boosted performance to 0.549/0.563 on liver/kidney segmentation (p<0.001/p<0.001). Using a pre-trained model and fine-tuning on minimal real data boosted performance to 0.472/0.476 (p=0.325/p<0.001), while using a pre-trained model and fine-tuning on minimal real data with synthetic gap augmentation yielded similar results (0.571/0.538; p<0.001/p<0.001) to solely training on minimal real data with synthetic gap augmentation without synthetic pretraining.

Training in low real data conditions (2 scans) achieved an average Dice score of 0.578/0.579 on liver/kidney segmentation. Synthetic gap augmentation boosted performance to 0.611/0.676 on liver/kidney segmentation (p=0.560/p<0.001). Using a pre-trained model and fine-tuning on low real data boosted performance to 0.584/0.613 (p=0.118/p<0.01), while using a pre-trained model and fine-tuning on low real data with synthetic gap augmentation yielded similar results (0.626/0.659; p=0.987/p<0.001) to solely training on low real data with synthetic gap augmentation without synthetic pretraining.

Training on moderate real data conditions (7 scans) achieved an average Dice score of 0.692/0.818 on liver/kidney segmentation. Synthetic gap augmentation yielded similar performance of 0.702/0.792 (p=0.640/p<0.01). Using a pre-trained model and fine-tuning on moderate real data led to decreased performance at 0.609/0.678 (p<0.001/p<0.001), while using a pre-trained model and fine-tuning on moderate data with synthetic gap augmentation also led to decreased performance (0.618/0.642; p<0.001/p<0.001).

Training on full real data conditions (all 14 scans) achieved an average Dice score of 0.750/0.845 on liver/kidney segmentation. Augmenting with 10%, 50%, and 100% synthetic data yielded similar performances of 0.766/0.849, 0.763/0.862, and 0.757/0.857, respectively (p <0.05/p=0.315, p=0.737/p=0.726, p<0.05/p=0.368). Using a pre-trained model and fine-tuning on full real data yielded a similar performance of 0.727/0.846 (p<0.001/p=0.309) while using a pre-trained model and fine-tuning on full real data with 10%, 50%, and 100% of synthetic data augmentation yielded decreased performance of 0.667/0.769, 0.644/0.711, and 0.682/0.753, respectively (p<0.001/p<0.001, p<0.001/p<0.001 , p<0.001/p<0.001).

*Labeled Lumbar Spine MRI*

The Labeled Lumbar Spine MRI dataset consists of 1,545 T2 MRI images. For each fold of the 5-fold cross-validation, the minimal/low/moderate/full real data condition was defined as 6/30/555/1113 of the MRI images; 123 images were used for validation, and 309 images were used for testing. Results are summarized in **Table 2c,** and corresponding statistics in **Tables S6a**, **S6b**, and **S6c**.

For the Labeled Lumbar Spine MRI dataset, training nnU-Net models in minimal real data conditions achieved an average Dice score of 0.924/0.772/0.852 on IVD/TS/PE segmentation. Synthetic gap augmentation in training yielded performance of 0.919/0.700/0.779 on IVD/TS/PE segmentation ($p<0.001/p<0.001/p<0.001$). Using a pre-trained model and fine-tuning on minimal real data boosted performance to 0.957/0.831/0.893 ($p<0.001/p<0.001/p<0.001$), while using a pre-trained model and fine-tuning on minimal real data with 99% synthetic data augmentation yielded similar results (0.924/0.715/0.788; $p<0.001/p<0.001/p<0.001$) to solely training on minimal real data with synthetic gap augmentation without pretraining.

Training in low real data conditions achieved an average Dice score of 0.967/0.857/0.907 on IVD/TS/PE segmentation. Synthetic gap augmentation in training yielded performance of 0.941/0.776/0.835 on IVD/TS/PE segmentation ($p<0.001/p<0.001/p<0.001$). Using a pre-trained model and fine-tuning on low real data yielded similar performance as without pretraining (0.970/0.867/0.912, $p<0.001/p<0.001/p<0.001$), while using a pre-trained model and fine-tuning on low real data with synthetic gap augmentation yielded similar results (0.945/0.789/0.843; $p<0.001/p<0.001/p<0.001$) to solely training on low real data with synthetic gap augmentation without pretraining.

Training on moderate real data conditions achieved an average Dice score of 0.974/0.899/0.927 on IVD/TS/PE segmentation. Synthetic gap augmentation in training yielded a similar performance of 0.973/0.893/0.924 ($p <0.001/p<0.001/p<0.001$). Using a pre-trained model and fine-tuning on moderate real data yielded similar performance of 0.975/0.901/0.928 ($p<0.001/p<0.001/p<0.001$), while using a pre-trained model and fine-tuning on moderate data with synthetic gap augmentation yielded similar results (0.973/0.892/0.924; $p<0.001/p<0.001/p<0.001$) to solely training on moderate data with synthetic gap augmentation without pretraining.

Training on full real data conditions achieved an average Dice score of 0.975/0.906/0.929 on IVD/TS/PE segmentation. Augmenting with 10%, 50%, and 100% synthetic data yielded similar performances of 0.975/0.905/0.928, 0.975/0.902/0.927, and 0.974/0.900/0.926, respectively (see **Table S6a-c** for associated p values). Using a pre-trained model and fine-tuning on full real data boosted performance to 0.976/0.907/0.929 ($p<0.001/p=0.207/p<0.05$) while using a pre-trained model and fine-tuning on full real data with 10%, 50%, and 100% of synthetic data augmentation yielded similar results (0.975/0.905/0.929, 0.975/0.902/0.927, 0.974/0.899/0.926) (see **Table S6a-c** for associated p values).

*CVC-ClinicDB*

The CVC-ClinicDB dataset consists of 612 still images from endoscopic colonoscopy videos with segmented polyps. The minimal/low/moderate/full real data condition was defined as 5(1%)/44(10%)/220(50%)/440(100%) of the MRI images; 49 images were used for validation, and 123 images were used for testing. Results are summarized in **Table 2d,** and corresponding statistics in **Table S7**.

For the CVC-ClinicDB dataset, training nnU-Net models in minimal real data conditions (1% of images) achieved an average Dice score of 0.377 on polyp segmentation. Synthetic gap augmentation in training significantly boosted performance to 0.685 on polyp segmentation ($p<0.001$). Using a pre-trained model and fine-tuning on minimal real data significantly boosted performance to 0.624 ($p<0.001$), while using a pre-trained model and fine-tuning on minimal real data with synthetic gap augmentation yielded similar results (0.697; $p<0.001$) to solely training on minimal real data with synthetic gap augmentation without synthetic pretraining.

Training in low real data conditions (10% of images) achieved an average Dice score of 0.656 on polyp segmentation. Synthetic gap augmentation in training boosted performance to 0.755 on polyp segmentation ($p<0.001$). Using a pre-trained model and fine-tuning on low real data boosted performance to 0.795 ($p<0.001$), while using a pre-trained model and fine-tuning on low real data with synthetic gap augmentation yielded similar results (0.762; $p<0.001$) to training on low real data with synthetic gap augmentation without synthetic pretraining.

Training on moderate real data conditions (50% of images) achieved an average Dice score of 0.788 on polyp segmentation. Synthetic gap augmentation boosted performance to 0.830 ($p<0.001$). Using a pretrained model and fine-tuning on moderate real data boosted performance to 0.837 ($p<0.001$), while using a pre-trained model and fine-tuning on moderate data with synthetic gap augmentation yielded similar results (0.837; $p<0.001$) to training on moderate data with synthetic gap augmentation without synthetic pretraining.

Training on full real data conditions (all 612 images) achieved an average Dice score of 0.798 on polyp segmentation. Augmenting with 10%, 50%, and 100% synthetic data yielded improved performance of 0.819, 0.832, and 0.830, respectively ($p<0.001$, $p<0.001$, $p<0.001$). Using a pre-trained model and fine-tuning on full real data boosted performance to 0.841 ($p<0.001$), while using a pre-trained model and fine-tuning on full real data with 10%, 50%, and 100% synthetic data augmentation yielded similar results 0.842, 0.845, 0.844, respectively ($p<0.001$, $p<0.001$, $p<0.001$).

**DISCUSSION**

In medical imaging, the generation of large and robust fully labeled imaging datasets is challenging due to the time and expertise required to manually annotate pixel-wise segmentation masks, as well as the limited public availability of medical imaging due to privacy and regulatory concerns. These data limitations are consistently documented in the literature and pose a significant challenge for training machine learning models for medical imaging (33,34). While pre-trained weights for classification tasks are more readily accessible due to their generic nature, pre-trained weights tailored for segmentation are inherently task-specific, further emphasizing the importance of developing robust labeled datasets to improve segmentation outcomes in medical imaging applications.

GANs have been studied for their potential use to generate synthetic medical images for diverse applications. One specific paradigm is the use of GANs to produce synthetic images for machine learning model training to enhance performance on downstream segmentation tasks. Work by Guo et al. in developing MedGAN (15) and Osuala et al. in developing medigan (35) illustrate the potential for this paradigm to synthetically augment real datasets and boost segmentation performance, particularly in low data availability conditions.

However, while these studies utilized GANs to generate unlabeled synthetic medical images of a single/limited class or modality, our proposed RadImageGAN paradigm extends these capabilities and enables the generation of synthetic images from a plethora of imaging modalities, sequences, and pathologies. In this study, we excluded the ultrasound modality from RadImageNet due to its variations in orientation compared to CT and MRI, which would have negatively affected the development of the RadImageGAN generator. In addition to radiologic imaging, we expanded our scheme to colonoscopy imaging to examine the potential of synthetic imaging for other medical imaging modalities. Because colonoscopy is a domain of medical imaging techniques fundamentally different from radiological imaging, we developed two separate generators, RadImageGAN-CT/MR and RadImageGAN-Gastro. The 2 generators we trained in this study are able to generate medical images from 130 pathological classes, 3 modalities, and 12 anatomical regions. Moreover, with only a modest number of manual annotations per task, BigDatasetGAN further enables the generation of accurate pixel-wise segmentation masks on these synthetic data (RadImageGAN-Labeled). Our proposed deep learning paradigm enables the use of synthetic labeled images for real downstream applications to boost segmentation performance with two methods: 1) first, as a method of synthetic data augmentation ("synthetic augmentation"), and 2) as synthetic source data for developing pre-trained weights for transfer learning ("synthetic pretraining").

**Figure 4** illustrates performance on the diverse downstream segmentation tasks assessed in this study under various data availability conditions and synthetic enhancements. Note that using synthetic augmentation and/or pretraining was able to boost performance on all segmentation tasks in nearly all data availability conditions relative to training on solely real data. There are several data situations in which synthetic enhancement is particularly effective, as well as a few data situations in which synthetic enhancement may lead to poorer performance.

Generally, we find that in minimal/low data condition scenarios, adding synthetic data and/or using synthetic pre-trained weights can boost segmentation performance significantly. Combining both synthetic methods may sometimes yield further improvements in sparse data conditions, but we generally observe that the greatest improvements come from applying a single synthetic method. It is possible that using both methods might cause overfitting in learning synthetic image features during training.

We also find that for tasks in which relatively high performance was already achieved in minimal/low data conditions with solely real data (e.g., segmentation of TS and PE in Labeled Lumbar Spine MRI), synthetic augmentation and synthetic pretraining + augmentation yielded poorer performance, while pretraining led to an improvement in performance. This suggests that synthetic augmentation may be detrimental for segmentation tasks in which real data have strong signals (e.g., strong edge correlation with masks) and synthetic data are unable to achieve similar signal strength. Pretrained weights developed based on synthetic data can boost performance in these situations by providing more ideal weight initializations that can be further refined with the strong signal from real data.

In cases of full data availability, we find that adding a moderate or full amount of synthetic data can significantly boost the performance in tasks that don't have relatively high performance solely with real data, such as in BTCV-Abdomen, CHAOS-MRI, and CVC-ClinicDB. Future work may investigate optimal proportions for additional synthetic augmentation in different types of tasks.

Our study sought to address the significant barrier of limited medical imaging data availability for machine learning applications, which hinders model performance and limits potential medical applications. With the increased implementation of our proposed RadImageGAN synthetic paradigm, we hope to provide an additional technique for machine learning model development. This paradigm could enhance model performance on many downstream tasks in conditions of both imaging data scarcity and abundance by applying RadImageGAN-CT/MR/Gastro or developing new synthetic generators to augment relevant real datasets. Training these synthetic generators is computationally expensive, but once trained, it can be easily applied to many diverse downstream tasks. RadImageGAN-CT/MR and RadImageGAN-Gastro can already be utilized to synthetically enhance many potential downstream tasks (see **Table S1**), and we envision the future development of generators with even more diverse classes, modalities, and pathologies for research use. In addition, our study also demonstrated a scheme for developing labeled multi-class synthetic generators of both synthetic images and their respective annotation masks using BigDatasetGAN to obtain an automatically labeled dataset by manually annotating a few synthetic images. This further extends the utility of RadImageGAN for segmentation tasks. The synthetic datasets produced by RadImageGAN-Labeled can contribute to future medical imaging research as a direct method of synthetic data augmentation or an indirect method of synthetic pre-trained weight development. This scheme of RadImageGAN utility can further expand the performance of future medical imaging research.

There are also several limitations of our proposed RadImageGAN paradigm. First, to boost a specific downstream application, the generators must already contain this specific image modality and anatomy. The generators produced in this study covered limited imaging modalities, anatomies, and pathologies. For example, there are no pathologic labels from cardiac MRI images in our generators, so they are unable to enhance performance on those tasks. Secondly, imaging data such as CT and MRI are usually volumetric, but our generators were trained on individual images and are only able to generate individual image slices, as opposed to complete volumes. Third, the masks used to train RadImageGAN-Labeled were labeled by one expert, and this might lead to potential biases, thus undermining the accuracy of the resulting masks. Fourth, within one pathological label, we have a mixture of different sequences, views, anatomies, and imaging protocols, which may affect performance. For example, in the lumbar spine segmentation task, we selected the spine canal stenosis label from RadImageGAN, but this label also contains cervical spine images, which might negatively affect the performance as a data augmentation method.

In our future work, we will continue to expand the multi-class variability of RadImageGAN to generate images of more pathological labels, modalities, and anatomies. Within each pathological label, we will provide a more detailed classification of the sequences, views, and contrasts to refine the selection of imaging for synthetic enhancement. We also plan to investigate techniques to enable the generation of high-resolution volumetric imaging data and assess the value of this paradigm. Lastly, we plan to investigate the use of unsupervised learning GAN (pseudo-labeling) techniques to generate masks and further reduce the already minimal manual labeling required.

In this study, we have demonstrated the advanced capabilities of RadImageGAN to proficiently produce high-quality synthetic images from various modalities such as CT, MRI, and colonoscopy video images across an unprecedented 130 classes. Complementing this, we have showcased the ability to apply BigDatasetGAN and automatically generate synthetic masks utilizing a weakly-supervised approach to synthetic data. The synthetic images and pre-trained weights sourced from RadImageGAN-Labeled synthetic images offer a significant enhancement to downstream segmentation applications. For data augmentation objectives, synthetic data achieves its peak performance at a 50%-100% ratio of synthetic to real data. We believe that this paradigm can be applied to many medical imaging tasks and enhance performance, particularly under low data availability conditions.

## METHODS

### Datasets

RadImageGAN was trained on the RadImageNet CT/MR database, a subset of the full RadImageNet database, consisting of 880,314 CT and MR images with 124 pathologic labels from 102,774 patients (10), to produce the RadImageGAN-CT/MR generator. All images were resized to 16x16, 32x32, 64x64, 128x128, 256x256, and 512x512 px resolutions for each respective model training stage. With the same training scheme, the RadImageGAN-Gastro generator was trained on a subset of the HyperKvasir dataset of colonoscopy images consisting of 6 pathological classes with sufficient samples for training (27), with a total of 5,713 images.

### StyleGAN-XL

We proposed the use of StyleGAN-XL to generate high-resolution 512x512 px synthetic medical images based on our large-scale, diverse RadImageNet database and the HyperKvasir database. StyleGAN-XL is a new state-of-the-art for large-scale multi-class image synthesis that applies pre-trained feature network and classifier guidance to StyleGAN3 (36), and thus achieves better image synthesis performance on large-scale and diverse datasets (25). StyleGAN-XL is a GAN-based approach that can generate high-resolution synthetic images by using a progressive, growing approach. More synthetic layers and super-resolution layers are added to the generator as the training progresses from low resolution to high resolution. In this study, we used 8 NVIDIA DGX1-A100 GPUs for StyleGAN-XL model training. For the development of RadImageGAN, we started with low-resolution outputs at 16x16 px and progressively trained to higher-resolution outputs at 512x512 px. For the initial 16x16 stage, the stem model was trained from scratch with 10 synthetic layers and 7 head layers. The Fréchet Inception Distance (FID) was used to monitor the training process, and the model with the lowest FID score was selected as the pre-trained weights for the next higher-resolution stage. For each resolution increase, the model utilizes the previous stage's model as pre-trained weights and adds 7 additional head layers. For the final 512x512 stage, only 5 head layers were added, giving 33 head layers in total. For the 16x16 and 32x32 stages, the batch size was set to 2048, and for all higher-resolution stages, the batch size was set to 256. X-axis flip data augmentation was enabled for all stages except the final 512x512 training stage.

### Downstream Datasets

**Table S8** presents a summary of the four downstream datasets by modality, image series, ROI targets, and dataset size. These datasets were chosen specifically to evaluate RadImageGAN's ability to interpret unseen medical imaging data sources. A brief description of each dataset is provided below:

1. BTCV-Abdomen (28): This dataset consists of 30 randomly selected abdomen CT scans with annotations from patients undergoing a colorectal cancer chemotherapy trial or a retrospective ventral hernia study. The scans were taken in the portal venous contrast phase and captured volumes of various sizes. Each scan corresponds with a segmentation map of 13 abdominal organs. For this study, liver and kidney segmentations were used.

2. CHAOS-MRI T2 (29): This consists of 20 MRI abdominal scans with annotations from the T2-SPIR sequence. This is a subset of the full combined CT-MRI dataset. Each dataset consists of 26 to 50 axial slices with a resolution of 256 x 256 pixels. In this study, only axial view images were utilized for evaluation. Each scan is associated with a labeled segmentation map of the kidneys, liver, and spleen. For this study, liver and kidney segmentations were used.

3. Labeled Lumbar Spine MRI (30): This dataset consists of 1,545 T2 MRI lumbar spine images obtained from an anonymized clinical study of 515 patients experiencing symptomatic back pain. Images were taken from either a sagittal or axial view and included the lowest three vertebrae and intervertebral discs (IVDs). The four ROIs were the Posterior Element (PE), Thecal Sac (TS), IVD, and the region between Anterior and Posterior (AAP) vertebrae elements. This dataset was chosen for its size and subsequent variability in age, gender, severity of illness, and imaging equipment. For this study, PE, TS, and IVD segmentations were used.

4. CVC-ClinicDB (31): The CVC-ClinicDB dataset consists of 612 still images sourced from endoscopic colonoscopy videos. Some images contain colorectal polyps. Each image corresponds to a segmentation map of the polyps.

For the BTCV-Abdomen, CHAOS-MRI, and CVC-ClinicDB datasets, all images were resized to 512x512 px. For the Labeled Lumbar Spine MRI dataset, all images were resized to 320x320 px.

**Data Validation**

For each downstream application, 5-fold cross-validation was conducted. In the BTCV-Abdomen and CHAOS-MRI, these two datasets were collected and resampled by patient, while the CVC-ClinicDB and Labeled Lumbar Spine MRI datasets were collected and resampled by individual image. The validation and test datasets were maintained for each fold of cross-validation.

**BigDatasetGAN**

We then proposed the use of BigDatasetGAN to generate high-resolution 512x512 px pixel-wise labeled synthetic medical images from RadImageGAN, turning RadImageGAN into a labeled dataset generator. BigDatasetGAN is a new state-of-the-art model extending DatasetGAN (37) that enables multi-class generation of pixel-wise annotated synthetic images from a class-conditional generative model using only a few manually annotated images per class (26). This is achieved by enhancing the pre-trained multi-class conditional generative model with a segmentation branch called a "feature interpreter" that is able to take a generator's class-specific features and class labels as input and output a set of all pixel-wise labels for each class. We follow the methods introduced by Li et al. and use a similar feature interpreter architecture for the RadImageGAN-Labeled generator.

In this study, 50 synthetic images were manually annotated by a senior radiologist (T.D.) for each downstream task on the BTCV-Abdomen, CHAOS-MRI, and Labeled Lumbar Spine MRI. For axial CT and MRI abdomen images, liver and kidney regions were annotated. For axial MRI spine images, IVD, TS, and PE regions were annotated. For colonoscopy images, polyps were annotated. 50 synthetic colonoscopy images on the CVC-ClinicDB dataset were manually annotated for polyps by a senior machine learning applied scientist (S.H.).

During the training of RadImageGAN-Labeled, we utilized hyperparameters of 0.0001 for learning rate, 4 for batch size, and 100 training epochs.

**Baseline model - nnU-Net**

In this study, a 2D nnU-Net (32) was used as the baseline model for training in all segmentation tasks and data conditions since its efficiency and competency in medical segmentation tasks have been well-established and it can automatically be adapted to any given dataset. For each downstream task, each data scenario was constructed as an individual dataset for nnU-Net model development. The default training parameters were automatically generated by the *nnU-Netv2_plan_and_preprocess* function, and each model was trained with 100 epochs using the planned parameters.

**Pretrained weights generated from synthetic data**

For each application, five-fold cross-validation was used to train five models from scratch using a full-size dataset of synthetic images and the corresponding validation and testing datasets from each fold. The model with the best validation performance was selected as the pre-trained weights for the specific downstream dataset and associated segmentation tasks.

**Statistics**

For each of the downstream datasets, Shapiro-Wilk tests were conducted to evaluate the normality of each distribution of Dice scores. Since every Shapiro-Wilk test yielded a p-value less than alpha = 0.05, Wilcoxon signed-rank tests were used across all pairs of columns in each downstream dataset to test for statistically significant differences in the mean Dice scores of paired observations. Both tests were performed with the Scipy package (version 1.10.1) using Python (version 3.10.9). The 95% CIs of dice scores for AI models were calculated by the bootstrap resampling with 1000 resamples.

**Data Availability**

In this study, the development of RadImageGAN utilized RadImageNet and HyperKvasir data. RadImageNet data is available by request via https://www.radimagenet.com/, and the full dataset of HyperKvasir can be downloaded via https://osf.io/mh9sj/. The four public segmentations datasets used in this study can be downloaded via the following links: BTCV-Abdomen (https://www.synapse.org/#!Synapse:syn3193805/wiki/217789), CHAOS-MRI T2 (https://chaos.grand-challenge.org/), Labeled Lumbar Spine MRI (https://data.mendeley.com/datasets/zbf6b4pttk/2), and CVC-ClinicDB (https://www.kaggle.com/datasets/balraj98/cvcclinicdb).

**Code Availability**

RadImageGAN generators and the model training process are available in the following GitHub repository: https://github.com/lzl199704/RadImageGAN.

**Acknowledgements**

X.M. was supported by the National Center for Advancing Translational Sciences (NCATS) TL1TR004420 NRSA TL1 Training Core in Transdisciplinary Clinical and Translational Science (CTSA).

**Competing interests**
T.D. is managing partner of RadImageNet LLC.

**Author Contributions**
Z.L., S.H., D.L., and X.M. developed the models and generators. Z.L. M.Y., A.Z., and X.M. created the figures. Z.L., A.Z., X.M., S.H., Z.A.F. and T.D. designed the experiments. S.H. and T.D. annotated the datasets. Z.L. and J.G. performed the statistical analysis. X.M., Z.A.F., T.D., and S.H. supervised the work. All authors wrote and reviewed the manuscript.

# FIGURES

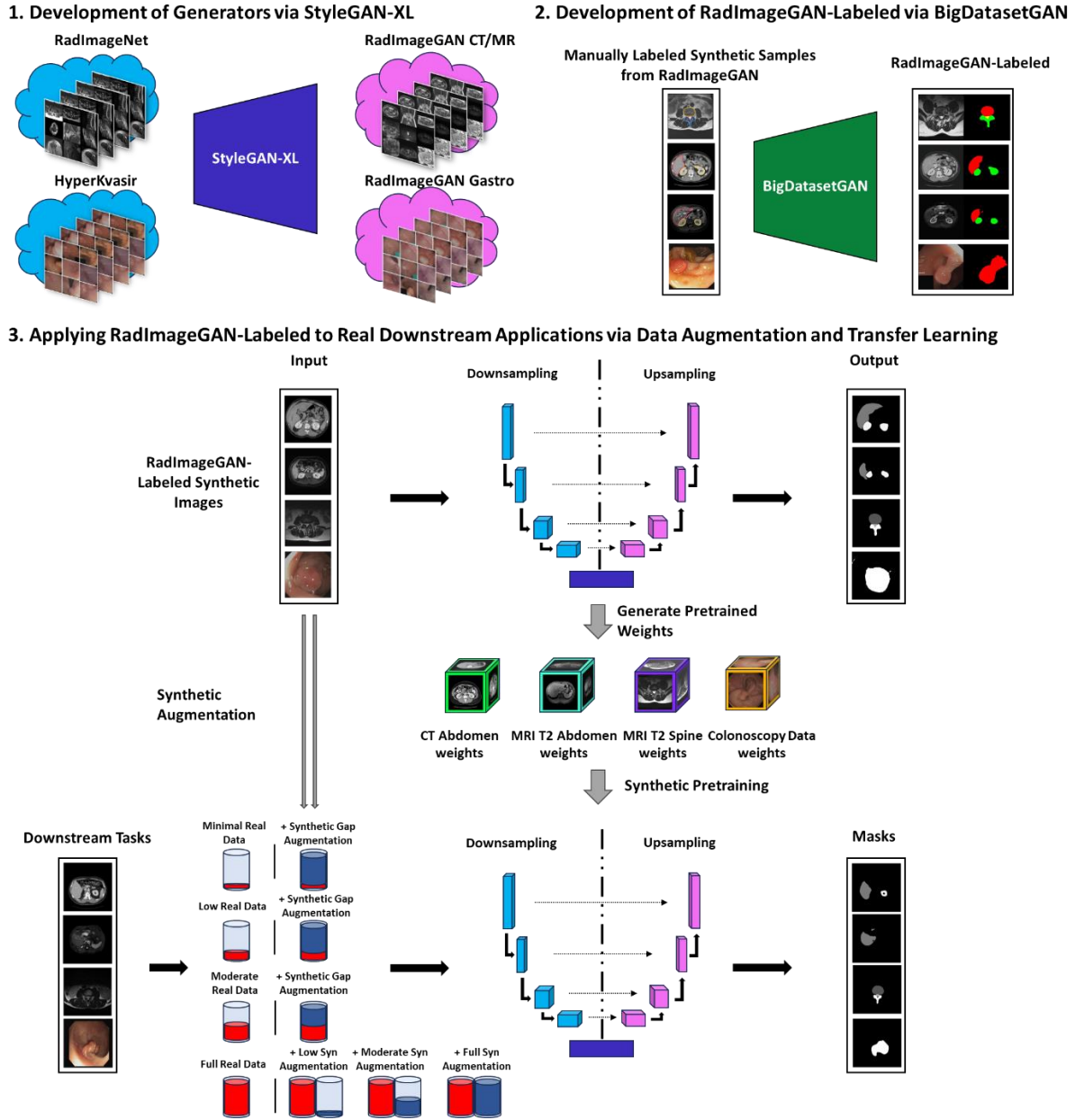

**Figure 1**: Overview of the study design. Step 1: RadImageGAN generators were developed by training StyleGAN-XL on a subset of the RadImageNet database and the HyperKvasir dataset of colonoscopy images; Step 2: By manually annotating 50 synthetic images for each specific downstream task, BigDatasetGAN was trained to develop RadImageGAN-Labeled for the generation of fully labeled synthetic images; Step 3: The impact of RadImageGAN-Labeled on segmentation performance is evaluated via synthetic data augmentation and/or transfer learning. Synthetic augmentation was evaluated under different data availability scenarios by taking increasingly larger subsets of the available training dataset (minimal, real, low, and full real datasets). Synthetic pretraining was evaluated using pretrained weights developed from full-size synthetic image datasets and fine-tuning on non-augmented and augmented datasets.

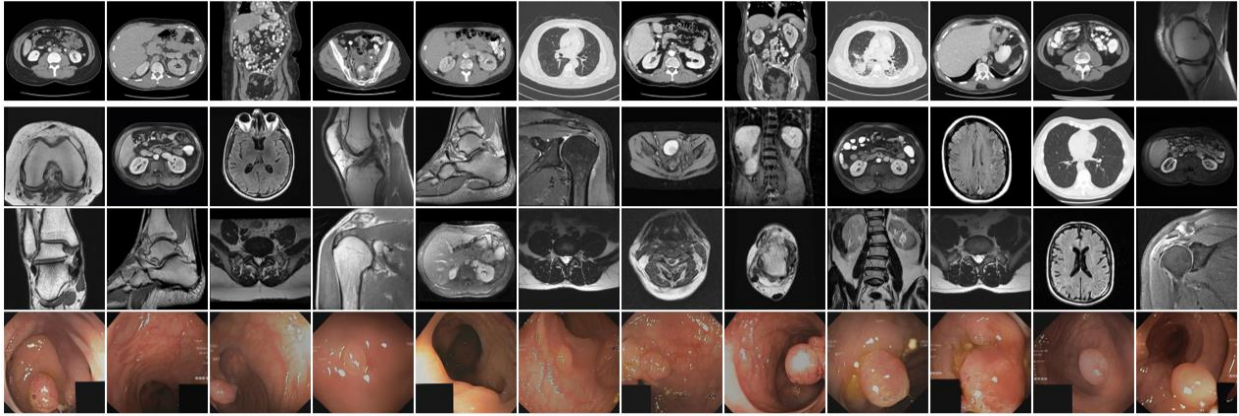

**Figure 2**: Representative images produced by the RadImageGAN generators across different sequences, views, and contrast.

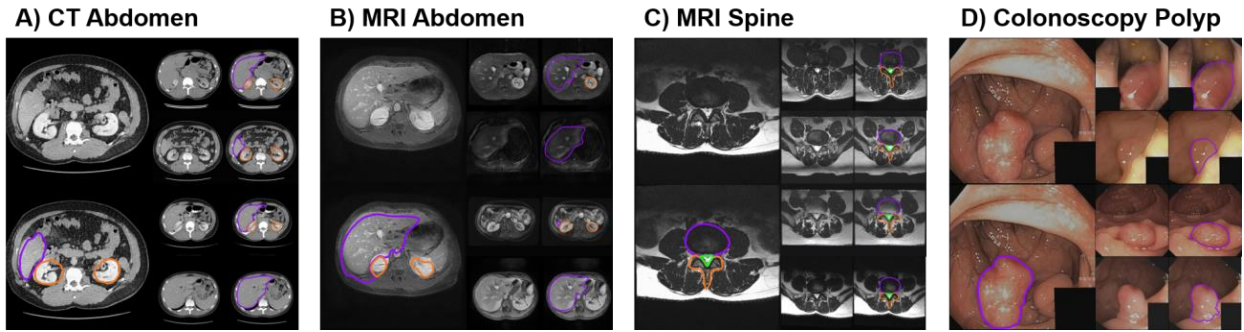

**Figure 3**. Representative images and masks produced by RadImageGAN-Labeled after training of BigDatasetGAN on minimal manual annotations of synthetic RadImageGAN images for each downstream dataset: A) CT Abdomen (for the BTCV-Abdomen downstream dataset; liver and kidney segmented); B) MRI Abdomen (for the CHAOS-MRI T2 downstream dataset; liver and kidney segmented); C) MRI Spine (for the Labeled Lumbar Spine MRI downstream dataset; intervertebral disc (IVD), posterior element (PE), and thecal sac (TS) segmented); D) Colonoscopy Polyps (for the CVC-ClinicDB dataset; polyps segmented).

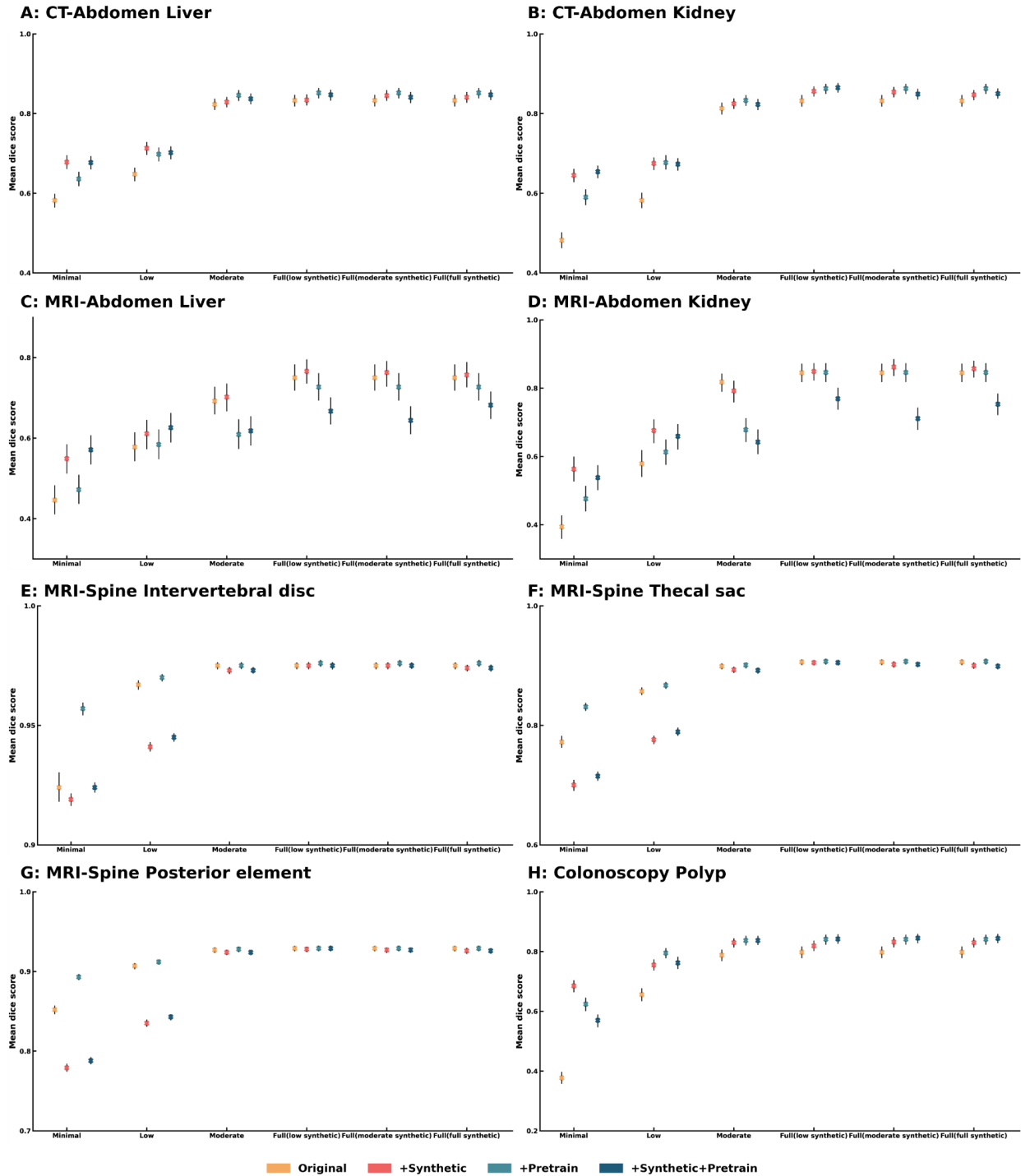

**Figure 4**. Bar chart comparisons of segmentation performance on downstream tasks in the four datasets without and with synthetic enhancing methods by task. Each task was grouped into 6 data conditions (minimal real data with synthetic gap augmentation, low real data with synthetic gap augmentation, moderate real data with synthetic gap augmentation, full real data with low synthetic augmentation, full real data with moderate synthetic augmentation, and full real data with full synthetic augmentation). Within each data condition, the four colors represent different

training situations of using only real data, using real data with synthetic augmentation, using real data with synthetic pretraining, and combining both synthetic methods. The black bar represents the 95% confidence interval of performance on each task: A) Liver segmentation on BTCV-Abdomen; B) Kidney segmentation on BTCV-Abdomen; C) Liver segmentation on CHAOS-MRI; D) Kidney segmentation on CHAOS-MRI; E) IVD segmentation on Labeled Lumbar Spine MRI; F) TS segmentation on Labeled Lumbar Spine MRI; G) PE segmentation on Labeled Lumbar Spine MRI; H) Polyp segmentation on CVC-ClinicDB.

# TABLES

**Table 1.** Summary of the 10 data conditions for synthetic augmentation experiments. Various proportions of real and synthetic imaging data for training were used to simulate conditions of real data abundance or scarcity and to understand the relative effect of training with datasets enhanced with synthetic images.

| Data Condition | Description | Dataset Selection Details |
|---|---|---|
| Minimal real data | Smallest possible imaging dataset subset for model training obtained from real patient imaging | 1 scan or 1% of images from a real dataset |
| Minimal real data + synthetic gap augmentation | Smallest possible real imaging dataset subset augmented by a synthetic imaging dataset such that overall augmented dataset is of a similar size to the original dataset | 1 scan or 1% real + 99% synthetic |
| Low real data | Small subset of an imaging dataset for model training obtained from real patient imaging | 2 scans or 10% of images from a real dataset |
| Low real data + synthetic gap augmentation | Small subset of a real imaging dataset augmented by a synthetic imaging dataset such that overall augmented dataset is of a similar size to the original dataset | 2 scan or 10% real + 90% synthetic |
| Moderate real data | ~50% subset of a real imaging dataset for model training obtained from real patient imaging | 50% of images/studies from a real dataset |
| Moderate real data + synthetic gap augmentation | ~50% subset of a real imaging dataset augmented by a synthetic imaging dataset such that overall augmented dataset is of a similar size to the original dataset | 50% real + 50% synthetic |
| Full real data | A full imaging dataset for model training obtained from real patient imaging | 100% real |
| Full real data + low synthetic augmentation | Full real imaging dataset augmented by a synthetic imaging dataset ~10% the size of a relevant publicly available dataset | 100% real +10% synthetic |
| Full real data + moderate synthetic augmentation | Full real imaging dataset augmented by a synthetic imaging dataset ~50% the size of a relevant publicly available dataset | 100% real + 50% synthetic |

| Full real data + full synthetic augmentation | Full real imaging dataset augmented by a synthetic imaging dataset the same size of a relevant publicly available dataset | 100% real + 100% synthetic |

**Table 2a.** Performance on downstream BTCV-Abdomen segmentation tasks. Values shown are average Dice scores. 95% confidence intervals of average Dice scores are shown in parentheses.

| | liver | kidney | Liver (with pretrain) | Kidney (with pretrain) |
|---|---|---|---|---|
| Minimal real data | 0.582 (0.564, 0.600) | 0.482 (0.462, 0.502) | 0.636 (0.618, 0.654) | 0.590 (0.570, 0.610) |
| Minimal real data+synthetic gap augmentation | 0.678 (0.661, 0.695) | 0.645 (0.628, 0.662) | 0.677 (0.660, 0.694) | 0.654 (0.638, 0.670) |
| Low real data | 0.648 (0.630, 0.666) | 0.582 (0.563, 0.601) | 0.698 (0.680, 0.716) | 0.677 (0.659, 0.695) |
| Low real data+synthetic gap augmentation | 0.713 (0.696, 0.730) | 0.675 (0.658, 0.692) | 0.702 (0.685, 0.719) | 0.673 (0.657, 0.689) |
| Moderate real data | 0.823 (0.809, 0.837) | 0.813 (0.798, 0.828) | 0.846 (0.832, 0.860) | 0.833 (0.820, 0.846) |
| Moderate real data+synthetic gap augmentation | 0.829 (0.816, 0.842) | 0.825 (0.812, 0.838) | 0.837 (0.823, 0.851) | 0.823 (0.809, 0.837) |
| Full real data | 0.833 (0.818, 0.848) | 0.832 (0.818, 0.846) | 0.852 (0.838, 0.866) | 0.863 (0.850, 0.876) |
| Full real data+ low synthetic augmentation | 0.834 (0.820, 0.848) | 0.856 (0.843, 0.869) | 0.847 (0.833, 0.861) | 0.865 (0.853, 0.877) |
| Full real data+ moderate synthetic augmentation | 0.845 (0.832, 0.858) | 0.854 (0.841, 0.867) | 0.841 (0.826, 0.856) | 0.849 (0.835, 0.863) |
| Full real data+ full synthetic augmentation | 0.841 (0.827, 0.855) | 0.847 (0.833, 0.861) | 0.847 (0.834, 0.860) | 0.850 (0.838, 0.862) |

**Table 2b.** Performance on downstream CHAOS-MRI segmentation tasks. Values shown are average Dice scores. 95% confidence intervals of average Dice scores are shown in parentheses.

|  | liver | kidney | liver (with pretrain) | Kidney (with pretrain) |
|---|---|---|---|---|
| Minimal real data | 0.446 (0.411, 0.481) | 0.394 (0.359, 0.429) | 0.472 (0.437, 0.507) | 0.476 (0.439, 0.513) |
| Minimal real data+synthetic gap augmentation | 0.549 (0.512, 0.586) | 0.563 (0.527, 0.599) | 0.571 (0.535, 0.607) | 0.538 (0.501, 0.575) |
| Low real data | 0.578 (0.543, 0.613) | 0.579 (0.540, 0.618) | 0.584 (0.548, 0.620) | 0.613 (0.576, 0.650) |
| Low real data+synthetic gap augmentation | 0.611 (0.572, 0.650) | 0.676 (0.639, 0.713) | 0.626 (0.589, 0.663) | 0.659 (0.620, 0.698) |
| Moderate real data | 0.692 (0.659, 0.725) | 0.818 (0.789, 0.847) | 0.609 (0.573, 0.645) | 0.678 (0.642, 0.714) |
| Moderate real data+synthetic gap augmentation | 0.702 (0.667, 0.737) | 0.792 (0.758, 0.826) | 0.618 (0.582, 0.654) | 0.642 (0.607, 0.677) |
| Full real data | 0.750 (0.718, 0.782) | 0.845 (0.818, 0.872) | 0.727 (0.693, 0.761) | 0.846 (0.818, 0.874) |
| Full real data+ low synthetic augmentation | 0.766 (0.735, 0.796) | 0.849 (0.822, 0.876) | 0.667 (0.634, 0.700) | 0.769 (0.737, 0.801) |
| Full real data+ moderate synthetic augmentation | 0.763 (0.728, 0.798) | 0.862 (0.836, 0.888) | 0.644 (0.610, 0.678) | 0.711 (0.678, 0.744) |
| Full real data+ full synthetic augmentation | 0.757 (0.726, 0.788) | 0.857 (0.831, 0.883) | 0.682 (0.648, 0.716) | 0.753 (0.721, 0.785) |

**Table 2c.** Performance on downstream Labeled Lumbar Spine MRI segmentation tasks. Values shown are average Dice scores. 95% confidence intervals of average Dice scores are shown in parentheses.

|  | Intervertebral Disc (IVD) | Thecal Sac (TS) | Posterior Element (PE) | IVD (with pretrain) | TS (with pretrain) | PE (with pretrain) |
|---|---|---|---|---|---|---|
| Minimal real data | 0.924 (0.918, 0.930) | 0.772 (0.762, 0.782) | 0.852 (0.846, 0.858) | 0.957 (0.954, 0.960) | 0.831 (0.824, 0.838) | 0.893 (0.889, 0.897) |
| Minimal real data+synthetic gap augmentation | 0.919 (0.916, 0.922) | 0.700 (0.690, 0.710) | 0.779 (0.774, 0.784) | 0.924 (0.922, 0.926) | 0.715 (0.707, 0.723) | 0.788 (0.784, 0.792) |
| Low real data | 0.967 (0.965, 0.969) | 0.857 (0.851, 0.863) | 0.907 (0.903, 0.911) | 0.970 (0.968, 0.972) | 0.867 (0.861, 0.873) | 0.912 (0.908, 0.916) |
| Low real data+synthetic gap augmentation | 0.941 (0.939, 0.943) | 0.776 (0.769, 0.783) | 0.835 (0.831, 0.839) | 0.945 (0.943, 0.947) | 0.789 (0.782, 0.796) | 0.843 (0.839, 0.847) |
| Moderate real data | 0.974 (0.973, 0.977) | 0.899 (0.894, 0.904) | 0.927 (0.923, 0.931) | 0.975 (0.974, 0.976) | 0.901 (0.895, 0.907) | 0.928 (0.924, 0.932) |
| Moderate real data+synthetic gap augmentation | 0.973 (0.971, 0.975) | 0.893 (0.888, 0.898) | 0.924 (0.921, 0.927) | 0.973 (0.972, 0.974) | 0.892 (0.886, 0.898) | 0.924 (0.920, 0.928) |
| Full real data | 0.975 (0.973, 0.977) | 0.906 (0.901, 0.911) | 0.929 (0.925, 0.933) | 0.976 (0.975, 0.977) | 0.907 (0.902, 0.912) | 0.929 (0.925, 0.933) |
| Full real data+ low synthetic augmentation | 0.975 (0.973, 0.977) | 0.905 (0.900, 0.910) | 0.928 (0.924, 0.932) | 0.975 (0.973, 0.977) | 0.905 (0.900, 0.910) | 0.929 (0.925, 0.933) |
| Full real data+ moderate synthetic augmentation | 0.975 (0.973, 0.977) | 0.902 (0.897, 0.907) | 0.927 (0.923, 0.931) | 0.975 (0.974, 0.976) | 0.902 (0.897, 0.907) | 0.927 (0.923, 0.931) |
| Full real data+ full synthetic augmentation | 0.974 (0.973, 0.975) | 0.900 (0.894, 0.906) | 0.926 (0.922, 0.930) | 0.974 (0.973, 0.975) | 0.899 (0.894, 0.904) | 0.926 (0.922, 0.930) |

**Table 2d.** Performance on downstream CVC-ClinicDB segmentation tasks. Values shown are average Dice scores. 95% confidence intervals of average Dice scores are shown in parentheses.

|  | polyp | Polyp (with pretrain) |
|---|---|---|
| Minimal real data | 0.377 (0.358, 0.396) | 0.624 (0.601, 0.647) |
| Minimal real data+synthetic gap augmentation | 0.685 (0.664, 0.706) | 0.697 (0.547, 0.593) |
| Low real data | 0.656 (0.634, 0.678) | 0.795 (0.778, 0.812) |
| Low real data+synthetic gap augmentation | 0.755 (0.736, 0.774) | 0.762 (0.742, 0.782) |
| Moderate real data | 0.788 (0.768, 0.808) | 0.837 (0.821, 0.853) |
| Moderate real data+synthetic gap augmentation | 0.830 (0.814, 0.846) | 0.837 (0.822, 0.852) |
| Full real data | 0.798 (0.778, 0.818) | 0.841 (0.823, 0.859) |
| Full real data+ low synthetic augmentation | 0.819 (0.802, 0.836) | 0.842 (0.826, 0.858) |
| Full real data+ moderate synthetic augmentation | 0.832 (0.814, 0.850) | 0.845 (0.828, 0.862) |
| Full real data+ full synthetic augmentation | 0.830 (0.814, 0.846) | 0.844 (0.829, 0.859) |

**SUPPLEMENTARY**

**Table S1.** Summary of RadImageGAN pathological classes and imaging modalities.

| Pathological Label | Class index | Imaging modality | Anatomy |
| --- | --- | --- | --- |
| abd_normal | 0 (CT/MR) | CT | Abdomen |
| af_normal | 1 | MRI | ankle/foot |
| ankle_accessory ossicle | 2 | MRI | Ankle |
| ankle_achilles tear | 3 | MRI | Ankle |
| ankle_achilles tendinosis | 4 | MRI | Ankle |
| ankle_atfl sprain | 5 | MRI | Ankle |
| ankle_atfl tear | 6 | MRI | Ankle |
| ankle_bursitis | 7 | MRI | Ankle |
| ankle_cfl sprain | 8 | MRI | Ankle |
| ankle_cfl tear | 9 | MRI | Ankle |
| ankle_fracture | 10 | MRI | Ankle |
| ankle_ganglion | 11 | MRI | Ankle |
| ankle_joint effusion | 12 | MRI | Ankle |
| ankle_marrow edema | 13 | MRI | Ankle |
| ankle_oa | 14 | MRI | Ankle |
| ankle_osteochondral lesion | 15 | MRI | Ankle |
| ankle_pb tear | 16 | MRI | Ankle |
| ankle_pl tendinosis | 17 | MRI | Ankle |
| ankle_plantar fasciitis | 18 | MRI | Ankle |
| ankle_post op | 19 | MRI | Ankle |
| ankle_ptt tendinosis | 20 | MRI | Ankle |
| ankle_soft tissue edema | 21 | MRI | Ankle |
| ankle_tenosynovitis | 22 | MRI | Ankle |
| brain-normal | 23 | MRI | Brain |
| brain_chronic infarct | 24 | MRI | Brain |
| brain_encephalomalacia | 25 | MRI | Brain |
| brain_intra-axial metastases | 26 | MRI | Brain |
| brain_meningioma | 27 | MRI | Brain |
| brain_microvascular disease | 28 | MRI | Brain |
| foot_accessory ossicle | 29 | MRI | Foot |
| foot_achilles tendinosis | 30 | MRI | Foot |
| foot_atfl sprain | 31 | MRI | Foot |
| foot_atfl tear | 32 | MRI | Foot |
| foot_bursitis | 33 | MRI | Foot |
| foot_cfl tear | 34 | MRI | Foot |
| foot_fracture | 35 | MRI | Foot |
| foot_ganglion | 36 | MRI | Foot |
| foot_joint effusion | 37 | MRI | Foot |
| foot_marrow edema | 38 | MRI | Foot |
| foot_neuroma | 39 | MRI | Foot |
| foot_oa | 40 | MRI | Foot |
| foot_osteochondral lesion | 41 | MRI | Foot |
| foot_pb tear | 42 | MRI | Foot |
| foot_pl tendinosis | 43 | MRI | Foot |
| foot_post op | 44 | MRI | Foot |
| foot_ptt tendinosis | 45 | MRI | Foot |
| foot_soft tissue edema | 46 | MRI | Foot |
| foot_soft tissue mass | 47 | MRI | Foot |
| foot_tenosynovitis | 48 | MRI | Foot |
| hip-normal | 49 | MRI | Hip |
| hip_bursitis | 50 | MRI | Hip |
| hip_cartilage abn nos | 51 | MRI | Hip |
| hip_labral tear | 52 | MRI | Hip |
| hip_marrow edema | 53 | MRI | Hip |
| hip_osseous lesion | 54 | MRI | Hip |
| hip_post op | 55 | MRI | Hip |
| hip_soft tissue edema | 56 | MRI | Hip |
| knee-chondral abnormality | 57 | MRI | Knee |
| knee-meniscal abnormality | 58 | MRI | Knee |
| knee-normal | 59 | MRI | Knee |
| knee_acl sprain | 60 | MRI | Knee |
| knee_acl tear | 61 | MRI | Knee |
| knee_bakers cyst | 62 | MRI | Knee |
| knee_chondral wear | 63 | MRI | Knee |
| knee_effusion | 64 | MRI | Knee |
| knee_fracture | 65 | MRI | Knee |
| knee_ganglion | 66 | MRI | Knee |
| knee_iliotibial band syndrome | 67 | MRI | Knee |
| knee_loose body | 68 | MRI | Knee |
| knee_mcl sprain | 69 | MRI | Knee |
| knee_mcl tear | 70 | MRI | Knee |
| knee_meniscal degeneration | 71 | MRI | Knee |
| knee_meniscal tear | 72 | MRI | Knee |
| knee_osseous contusion | 73 | MRI | Knee |
| knee_osteoarthritis | 74 | MRI | Knee |
| knee_parameniscal cyst | 75 | MRI | Knee |
| knee_patellar tendinosis | 76 | MRI | Knee |
| knee_synovitis | 77 | MRI | Knee |
| lung-normal | 78 | CT | Lung |
| lung_abd_Bronchiectasis | 79 | CT | Lung |
| lung_abd_Emphysema | 80 | CT | Lung |
| lung_abd_Pneumonia | 81 | CT | Lung |
| lung_abd_Scar/atelectasis | 82 | CT | Lung |
| lung_abd_ascites | 83 | CT | Abdomen |
| lung_abd_diverticulitis | 84 | CT | Abdomen |
| lung_abd_diverticulosis | 85 | CT | Abdomen |
| lung_abd_gallstone | 86 | CT | Abdomen |
| lung_abd_hemangioma nos | 87 | CT | Abdomen |
| lung_abd_hepatic cyst | 88 | CT | Abdomen |
| lung_abd_hepatic lesion nos | 89 | CT | Abdomen |
| lung_abd_hiatal hernia | 90 | CT | Abdomen |
| lung_abd_hydronephrosis | 91 | CT | Abdomen |
| lung_abd_inflammation nos | 92 | CT | Abdomen |
| lung_abd_mass nos | 93 | CT | Abdomen |
| lung_abd_myoma | 94 | CT | Abdomen |
| lung_abd_ovarian/adnexal cyst | 95 | CT | Abdomen |
| lung_abd_post op | 96 | CT | Abdomen |
| lung_abd_pulm nodule | 97 | CT | Lung |
| lung_abd_renal cyst | 98 | CT | Abdomen |
| lung_abd_renal lesion nos | 99 | CT | Abdomen |
| lung_abd_sclerotic osseous lesion nos | 100 | CT | Abdomen |
| lung_abd_ureteral calculus | 101 | CT | Abdomen |
| mri_abd_hepatic cyst | 102 | MRI | Abdomen |
| mri_abd_hepatic lesion nos | 103 | MRI | Abdomen |
| mri_abd_myoma | 104 | MRI | Abdomen |
| mri_abd_ovarian/adnexal cyst | 105 | MRI | Abdomen |
| mri_abd_renal cyst | 106 | MRI | Abdomen |
| mriabd-normal | 107 | MRI | Abdomen |
| shoulder-normal | 108 | MRI | Shoulder |
| shoulder_acj oa | 109 | MRI | Shoulder |
| shoulder_biceps tendinosis | 110 | MRI | Shoulder |
| shoulder_bursitis | 111 | MRI | Shoulder |
| shoulder_capsulitis | 112 | MRI | Shoulder |
| shoulder_ghj oa | 113 | MRI | Shoulder |
| shoulder_labral tear | 114 | MRI | Shoulder |
| shoulder_post op | 115 | MRI | Shoulder |
| shoulder_supraspinatus tear | 116 | MRI | Shoulder |
| shoulder_supraspinatus tendinosis | 117 | MRI | Shoulder |
| spine-normal | 118 | MRI | Spine |
| spine_canal stenosis | 119 | MRI | Spine |
| spine_ddd | 120 | MRI | Spine |
| spine_foraminal stenosis | 121 | MRI | Spine |
| spine_herniation | 122 | MRI | Spine |
| spine_scoliosis | 123 | MRI | Spine |

| dyed_lifted_polyps | 0 (Gastro) | Colonoscopy | N/A |
| dyed_resection_margins | 1 | Colonoscopy | N/A |
| polyps | 2 | Colonoscopy | N/A |
| pylorus | 3 | Colonoscopy | N/A |
| retroflex_stomach | 4 | Colonoscopy | N/A |
| z_line | 5 | Colonoscopy | N/A |

**Table S2.** Summary of synthetic generator performance quantified by average FID and computational training time required at each training stage for RadImageGAN-CT/MR.

| Output Resolution | Pretrained network | FID | Time |
|---|---|---|---|
| 16x16 | None | 6.40 | 44h |
| 32x32 | 16x16 | 7.75 | 53h |
| 64x64 | 32x32 | 7.60 | 62h |
| 128x128 | 64x64 | 8.85 | 109h |
| 256x256 | 128x128 | 9.23 | 120h |
| 512x512 | 256x256 | 9.37 | 183h |
|  |  | Total | 4,563 A100 hours |

**Table S3.** Summary of synthetic generator performance quantified by average FID and computational training time required at each training stage for RadImageGAN-Gastro.

| Output Resolution | Pretrained network | FID | Time |
|---|---|---|---|
| 16x16 | None | 8.08 | 20hr |
| 32x32 | 16x16 | 9.12 | 17h |
| 64x64 | 32x32 | 11.36 | 16h |
| 128x128 | 64x64 | 8.74 | 74h |
| 256x256 | 128x128 | 5.50 | 91h |
| 512x512 | 256x256 | 5.05 | 168h |
|  |  | Total | 3,088 V100 hours |

**Table S4a.** P values for effects on liver segmentation performance in the BTCV-Abdomen dataset under each synthetic training method condition.

|  |  | Minimal real data | Low real data | Moderate real data | Full real data |
|---|---|---|---|---|---|
| + | Synthetic gap /full synthetic augmentation | <0.001 | 0.97147 | <0.001 | <0.001 |
| + | pretrain | <0.001 | <0.001 | <0.001 | <0.001 |
| +<br>+ | Synthetic gap /full Synthetic<br>pretrain | <0.001 | 0.46839 | <0.001 | <0.001 |
| + | low synthetic augmentation | N/A | N/A | N/A | <0.001 |
| +<br>+ | low synthetic augmentation<br>pretrain | N/A | N/A | N/A | 0.10395 |
| + | moderate synthetic augmentation | N/A | N/A | N/A | <0.001 |
| +<br>+ | moderate synthetic augmentation<br>pretrain | N/A | N/A | N/A | <0.001 |

**Table S4b.** P values for effects on kidney segmentation performance in the BTCV-Abdomen dataset under each synthetic training method condition.

|  |  | Minimal real data | Low real data | Moderate real data | Full real data |
|---|---|---:|---:|---:|---:|
| + | Synthetic gap /full synthetic augmentation | <0.001 | <0.001 | <0.001 | <0.01 |
| + | pretrain | <0.001 | <0.001 | <0.001 | <0.001 |
| +<br>+ | Synthetic gap /full Synthetic<br>pretrain | <0.001 | <0.001 | <0.001 | <0.01 |
| + | low synthetic augmentation | N/A | N/A | N/A | 0.70967 |
| +<br>+ | low synthetic augmentation<br>pretrain | N/A | N/A | N/A | 0.13339 |
| + | moderate synthetic augmentation | N/A | N/A | N/A | 0.06103 |
| +<br>+ | moderate synthetic augmentation<br>pretrain | N/A | N/A | N/A | 0.3792 |

**Table S5a.** P values for effects on liver segmentation performance in the CHAOS-MRI dataset under each synthetic training method condition.

|  |  | Minimal real data | Low real data | Moderate real data | Full real data |
|---|---|---|---|---|---|
| + | Synthetic gap /full synthetic augmentation | <0.001 | 0.56019 | 0.64016 | <0.05 |
| + | pretrain | 0.32463 | 0.11804 | <0.001 | <0.001 |
| +<br>+ | Full Synthetic<br>pretrain | <0.001 | 0.98709 | <0.001 | <0.001 |
| + | low synthetic augmentation | N/A | N/A | N/A | <0.05 |
| +<br>+ | low synthetic augmentation<br>pretrain | N/A | N/A | N/A | <0.001 |
| + | moderate synthetic augmentation | N/A | N/A | N/A | 0.73672 |
| +<br>+ | moderate synthetic augmentation<br>pretrain | N/A | N/A | N/A | <0.001 |

**Table S5b.** P values for effects on kidney segmentation performance in the CHAOS-MRI dataset under each synthetic training method condition.

|  |  | Minimal real data | Low real data | Moderate real data | Full real data |
|---|---|---|---|---|---|
| + | Synthetic gap /full synthetic augmentation | <0.001 | <0.001 | <0.01 | 0.36823 |
| + | pretrain | <0.001 | <0.01 | <0.001 | 0.30913 |
| + + | Full Synthetic pretrain | <0.001 | <0.001 | <0.001 | <0.001 |
| + | low synthetic augmentation | N/A | N/A | N/A | 0.31536 |
| + + | low synthetic augmentation pretrain | N/A | N/A | N/A | <0.001 |
| + | moderate synthetic augmentation | N/A | N/A | N/A | 0.72617 |
| + + | moderate synthetic augmentation pretrain | N/A | N/A | N/A | <0.001 |

**Table S6a.** P values for effects on IVD segmentation performance in the Labeled Lumbar Spine MRI dataset under each synthetic training method condition.

|  | Minimal real data | Low real data | Moderate real data | Full real data |
|---|---|---|---|---|
| + Synthetic gap /full synthetic augmentation | <0.001 | <0.001 | <0.001 | <0.001 |
| + pretrain | <0.001 | <0.001 | <0.001 | <0.001 |
| + Full Synthetic + pretrain | <0.001 | <0.001 | <0.001 | <0.001 |
| + low synthetic augmentation | N/A | N/A | N/A | <0.05 |
| + low synthetic augmentation + pretrain | N/A | N/A | N/A | 0.11607 |
| + moderate synthetic augmentation | N/A | N/A | N/A | <0.001 |
| + moderate synthetic augmentation + pretrain | N/A | N/A | N/A | <0.001 |

**Table S6b.** P values for effects on TS segmentation performance in the Labeled Lumbar Spine MRI dataset under each synthetic training method condition.

|  | Minimal real data | Low real data | Moderate real data | Full real data |
|---|---|---|---|---|
| + Synthetic gap /full synthetic augmentation | <0.001 | <0.001 | <0.001 | <0.001 |
| + pretrain | <0.001 | <0.001 | <0.001 | 0.20726 |
| + Full Synthetic + pretrain | <0.001 | <0.001 | <0.001 | <0.001 |
| + low synthetic augmentation | N/A | N/A | N/A | <0.001 |
| + low synthetic augmentation + pretrain | N/A | N/A | N/A | <0.01 |
| + moderate synthetic augmentation | N/A | N/A | N/A | <0.001 |
| + moderate synthetic augmentation + pretrain | N/A | N/A | N/A | <0.001 |

**Table S6c.** P values for effects on PE segmentation performance in the Labeled Lumbar Spine MRI dataset under each synthetic training method condition.

|  | Minimal real data | Low real data | Moderate real data | Full real data |
|---|---|---|---|---|
| + Synthetic gap /full synthetic augmentation | <0.001 | <0.001 | <0.001 | <0.001 |
| + pretrain | <0.001 | <0.001 | <0.001 | <0.05 |
| + Full Synthetic + pretrain | <0.001 | <0.001 | <0.001 | <0.001 |
| + low synthetic augmentation | N/A | N/A | N/A | <0.001 |
| + low synthetic augmentation + pretrain | N/A | N/A | N/A | 0.11752 |
| + moderate synthetic augmentation | N/A | N/A | N/A | <0.001 |
| + moderate synthetic augmentation + pretrain | N/A | N/A | N/A | <0.001 |

**Table S7.** P values for effects on polyp segmentation performance in the CVC-ClinicDB dataset under each synthetic training method condition.

|  |  | Minimal real data | Low real data | Moderate real data | Full real data |
|---|---|---|---|---|---|
| + | Synthetic gap /full synthetic augmentation | <0.001 | <0.001 | <0.001 | <0.001 |
| + | pretrain | <0.001 | <0.001 | <0.001 | <0.001 |
| +<br>+ | Full Synthetic<br>pretrain | <0.001 | <0.001 | <0.001 | <0.001 |
| + | low synthetic augmentation | N/A | N/A | N/A | <0.001 |
| +<br>+ | low synthetic augmentation<br>pretrain | N/A | N/A | N/A | <0.001 |
| + | moderate synthetic augmentation | N/A | N/A | N/A | <0.001 |
| +<br>+ | moderate synthetic augmentation<br>pretrain | N/A | N/A | N/A | <0.001 |

**Table S8**. Downstream Task Summary

| Downstream Dataset | Total training set size (scans/images) | Modality | Number of labels used in the study | RadImageGAN label |
|---|---|---|---|---|
| BTCV-Abdomen | 30 scans | CT | 2 (liver, kidney) | 0 - abd_normal |
| CHAOS-MRI | 20 scans | MRI T2 | 2 (liver, kidney) | 107 - mri_abd_normal |
| Labeled Lumbar Spine MRI | 1545 images | MRI T2 | 3 (IVD, PE, TS) | 119 - spine_canal stenosis |
| CVC-ClinicDB | 612 images | Colonoscopy | 1 (polyp) | 2 - Polyp |